# On the classification of triggers of dynamic processes in geophysics

A.V. Guglielmi, A.D. Zavyalov, O.D. Zotov

*Schmidt Institute of Physics of the Earth RAS, Moscow, Russia*

*guglielmi@mail.ru, zavyalov@ifz.ru, ozotov@inbox.ru*

## Abstract

In geophysics, the problem of classifying triggers of dynamic processes in the lithosphere, hydrosphere, atmosphere, ionosphere and magnetosphere has arisen and needs to be solved. Triggers that induce catastrophic events, such as destructive earthquakes, require special attention. Classification is necessary in order to express the diversity of triggers in a limited number of ordered and well-identifiable species. This paper proposes a project for collective work on the systematization of triggers, i.e., the formation of a unified classification and nomenclature. A simplified (basic) and extended classification table are proposed. The basic matrix uses the principle of binary opposition. The matrix contains three categories of descending rank: type (natural, artificial), class (endogenous, exogenous) and specie (periodic, aperiodic). Three additional categories have been added to the expanded classification table. This results in 128 trigger varieties. The heuristic significance of the classification of triggers is emphasized. New phenomena discovered during the classification process are indicated: excitation of a strong aftershock by a round-the-world seismic echo, modulation of global seismicity by spheroidal oscillations of the Earth, the weekend effect in earthquake activity. In the magnetosphere, a cause-and-effect chain of triggers has been identified, called a trigger cascade. The existence of so-called antitriggers is indicated, which do not excite, as usually happens, but interrupt the ongoing dynamic process.

## 1. Introduction

A trigger is usually defined as a disturbance to a dynamic system that results in a significant change in its structure and/or functioning. There is a great diversity of triggers operating in the lithosphere, hydrosphere, atmosphere, ionosphere and magnetosphere of the Earth. Triggers that induce catastrophic events, in particular destructive earthquakes, attract special attention in geophysics. A complex problem of trigger classification has arisen in the geophysical community and requires a solution, as well as, perhaps, an equally complex problem of improving the taxonomy of catastrophic processes stimulated by one or another trigger.

At the 8th International Conference on Triggers in Geospheres, which was held from 18 to 22 May 2026 in Kolomna (*https://conf2026.idg.ras.ru*), a proposal was put forward to begin collective work on the taxonomy of triggers in geospheres, i.e., on the formation of a generally accepted classification and nomenclature [*Guglielmi et al.*, 2026a, b; *Zavyalov et al.*, 2026; *Zotov et al.*, 2026], and a decision was made to create a Working Group to coordinate efforts in this direction. Classification is necessary in order to express the diversity of triggers in a limited number of ordered and well-identifiable types. The need for classification arose also for the important reason that, thanks to the productive work of eight conferences on triggers in the geosphere, a new direction in geophysics has actually emerged, which can be conditionally called the study of triggers, or triggerology for short. The classification will facilitate the unification of knowledge about triggers and more successful coordination of research into trigger phenomena.

In this paper we will present one possible approach to the classification problem. We will consider a specific case of classification of earthquake triggers and try to find convincing arguments in favor of the idea that the systematics of seismic activity triggers can be used as a basis for developing a system of magnetospheric triggers, and probably also triggers in other geospheres.

## 2. Earthquake triggers

Let us sketch a classification of earthquake triggers as a model that can be used in the general classification of triggers in geophysical environments. Let us first introduce a simplified (basic) classification containing three categories of descending rank. Let's assign the names type, class and specie to the categories. We use the principle of binary opposition, dividing each category into two disjoint subsets that exhaust all triggers of a given category. Each of the two types is divided into two classes, and each class into two species. Thus, we get two types, four classes and eight specits of triggers. We will choose one specific basis to divide triggers at each classification level. When choosing a basis (feature, property of a trigger), we tried to ensure that it was logical and, if possible, did not cause difficulties when assigning each specific trigger to one or another subset. After analyzing many options, we selected the basic classification presented in Table 1.

**Table 1.** Basic classification of triggers

| **Type** | |
|---|---|
| Natural | Artificial |
| **Class** | |
| Endogenous | Exogenous |
| **Specie** | |
| Periodic | Aperiodic |

The basic matrix is designed to be suitable for systematizing triggers in any of the geospheres. In this case, the division of triggers into endogenous and exogenous is made according to the following criterion. If a trigger is excited in the same geosphere in which the dynamic system it affects is located, then we will call such a trigger endogenous. If a trigger originates in another geosphere, or even outside the Earth's magnetosphere, then such a trigger is called exogenous. A striking example of an exogenous trigger is a solar flare, which modifies dynamic processes in the ionosphere.

It is convenient to encode trigger species using three-digit numbers. To do this, we will write the number 1 in the cells in the left column of the table, and write the number 2 in the cells in the right column. Then, for example, the solar flare mentioned above would be classified as specie 122.

We consider Table 1 as a seed matrix, which can be supplemented with categories of one rank or another. Let us introduce a category of lower rank than species and call it variety. Let us choose a basis for dividing species into varieties that is of significant interest from the point of view of earthquake physics. Namely, we will try to characterize in the most general terms the mechanism of the trigger's effect on the source. In macroscopic physics, interaction can be forceful (mechanical, electromagnetic) or non-forceful (thermal, chemical). Thus, each species can have four varieties, listed in Table 2. As a result, we have 32 possible varieties of triggers.

From a theoretical point of view, it would be natural to also introduce an additional category (genus), of a higher rank than type, and divide it into two subsets that differ in the property of the trigger to be an additive or multiplicative disturbance of the earthquake source. In that case we would have 64 varieties of triggers. But the property of additivity or multiplicativity makes sense when looking at the earthquake source as a dynamic system in the light of the mathematical theory of catastrophes [*Guglielmi*, 2015]. In the widespread practical use of the proposed taxonomy, difficulties will most likely arise in deciding whether a given particular trigger is additive or multiplicative. However, it is useful to keep this category in mind and use it in cases where the identification of the trigger is not in doubt.

There is a logical basis for introducing a category intermediate between genus and type. The point is this. Typically, a trigger is considered to be, explicitly or implicitly, a disturbance that initiates a certain dynamic process, for example, the process of destruction of rocks in the source with the formation of a main discontinuity that excites the main shock of an earthquake. But a trigger, or rather an anti-trigger, can naturally also be called a disturbance that interrupts the current dynamic process. In the next section of the article we will point out the existence of antitriggers in the lithosphere and in the magnetosphere. We will call the additional category a family and divide it into two subsets – triggers and antitriggers.

The above considerations lead us to an extended classification of triggers, presented in Table 2. The table potentially contains 128 trigger varieties, divided into six categories.

**Table 2.** Extended classification of triggers

<table>
<tr><th colspan="4">Genus</th></tr>
<tr><td colspan="2">Additive</td><td colspan="2">Multiplicative</td></tr>
<tr><th colspan="4">Family</th></tr>
<tr><td colspan="2">Trigger</td><td colspan="2">Antitrigger</td></tr>
<tr><th colspan="4">Type</th></tr>
<tr><td colspan="2">Natural</td><td colspan="2">Artificial</td></tr>
<tr><th colspan="4">Class</th></tr>
<tr><td colspan="2">Endogenous</td><td colspan="2">Exogenous</td></tr>
<tr><th colspan="4">Specie</th></tr>
<tr><td colspan="2">Periodic</td><td colspan="2">Aperiodic</td></tr>
<tr><th colspan="4">Variety</th></tr>
<tr><td>Mechanical</td><td>Electromagnetic</td><td>Thermal</td><td>Chemical</td></tr>
</table>

## 3. Discussion

The basic matrix presented in Table 1 can be used without changes in developing a classification of triggers of electrodynamic processes in the magnetosphere. Let us give some typical examples.

The reconnection of magnetic field lines in the neutral layer of the geomagnetic tail is a specie 112 trigger. The trigger causes the injection of energetic charged particles into the auroral zones and the development of a complex of processes called a magnetospheric substorm. The interplanetary shock wave is a type 122 trigger. When the solar wind velocity is increased and the interplanetary magnetic field is in a favorable orientation, the trigger induces the sudden onset of a geomagnetic storm. We will also point out a strictly periodic anthropogenic trigger of unclear etiology, but of a very specific type 221, causing a reaction in the outer radiation belt and manifesting itself in the form of the Big Ben effect. The effect is that following the emission of a radio pulse of precise time into the ether, excitation of ultra-low-frequency electromagnetic oscillations of the radiation belt in the Pc1 range (0.2 – 5 Hz) is observed.

The trigger cascade in the form of a chain of triggers 122 ➡ 112 ➡ 112 is interesting. Initially, at the subsolar point of the magnetopause, a reconnection of geomagnetic lines and interplanetary magnetic field lines occurs. This first trigger leads to the transfer of geomagnetic lines from the periphery of the dayside magnetosphere into the geomagnetic tail. This results in a second trigger, the reconnection of geomagnetic lines in the neutral layer of the tail, which leads to the injection of charged particles into auroral zones (the third trigger) and a magnetospheric substorm.

We now want to discuss the heuristic potential of trigger classification. We will demonstrate, using specific examples, how classification helps in the search for and discovery of previously unknown phenomena.

A theoretical analysis of the question of the possible existence of a periodic endogenous natural earthquake trigger (code 111) led to the experimental discovery of modulation of global seismic activity by spheroidal oscillations of the Earth 0S2 with a period of 54 minutes. A previously unknown trigger with code 112 (aperiodic endogenous natural) was theoretically predicted and experimentally discovered in the aftershock activity. Trigger 112 has the appearance of a round-the-world seismic echo generated by the main shock of an earthquake. Returning to the epicenter approximately 3 hours after the main shock, the echo can induce an anomalous aftershock, the energy of which is tens of times higher than the energy of any other aftershock. Experimental evidence has been found that the cumulative effect of a round-the-world echo generated by a strong foreshock can induce the main shock of an earthquake. The weekend effect in global seismicity (type 221) deserves special mention. This calendar effect is caused by a trigger of as yet unclear origin.

Let us move on to a discussion of Table 2. In an extended classification, we must indicate the mechanism of action of a specific trigger on a specific dynamic process. Finding mechanisms is one of the central tasks of the theory of triggers, but experience shows that in many cases the problem does not lend itself to a simple solution. Sometimes we know about the existence of a trigger implicitly, without understanding its physical embodiment until a certain time. A typical example is the above-mentioned calendar trigger type 221, the existence of which we judge by the weekend effect on earthquake activity. In other cases, the physical nature of the trigger is known to us, but the mechanism of its action on the object remains uncertain. An example would be the excitation of a strong electrical impulse in the earth's crust from an MHD generator, followed by an increase in seismicity in the region. Thus, it is not possible to indicate the variety of trigger with certainty in all cases. On the other hand, in a number of cases the variety can be determined quite reliably. For example, when a large reservoir fills, an increase in regional seismicity is sometimes observed. In this case, we are dealing with a mechanical type of trigger 222. It is advisable to indicate the variety of trigger in those cases where this is possible.

In conclusion of the discussion, we note that the classification of disturbances affecting the dynamics of geophysical systems led us to the idea of identifying antitriggers as a special family. A pilot search was conducted and antitriggers were discovered in the magnetosphere and lithosphere. For example, based on the theory of permanent oscillations Pc3 of the magnetosphere [*Guglielmi*, 2007], one can confidently say that the boundaries between sectors of the interplanetary magnetic field are antitriggers in relation to Pc3. Moreover, in this particular case, it is possible to definitely indicate not only the family and species, but also the genus (multiplicative) and variety (electromagnetic) of the disturbance that interrupts the process of excitation of Pc3 when the Earth crosses the boundary between sectors of the interplanetary magnetic field.

## 4. Conclusion

We proposed a project for classifying triggers in geospheres in order to bring order and clarity to triggerology. Classification will give unity to the doctrine of triggers. In pursuance of the recommendation of the 8th Trigger Conference, complex collective work is now required to systematize triggers, i.e., to select classification tables that may differ from each other for triggers operating in a particular geosphere, as well as the particularly important work of assigning standardized names to triggers. The experience of a broad discussion in the geophysical community will show how productive the idea of the need to classify triggers of dynamic processes in the geospheres is.

***Acknowledgments.*** We sincerely thank G.G. Kocharyan for his support of the project to systematize the triggers of dynamic processes in geophysical environments. We express our deep gratitude to B.V. Dovbnya, F.Z. Feygin and A.S. Potapov, who made significant contributions to the study and classification of triggering phenomena in the lithosphere and magnetosphere. This work was supported by the Ministry of Science and Higher Education of the Russian Federation within the framework of state assignments of the Schmidt Institute of Physics of the Earth of the Russian Academy of Sciences.

This paper will be published in the journal "Dynamic Processes in Geospheres".